\documentclass{ifacconf}

\usepackage{graphicx}      
\usepackage{natbib}        
\usepackage{graphicx}      
\usepackage{tikz}
\usepackage{xcolor}
\usetikzlibrary{shapes,arrows,fit,positioning}
\usepackage{natbib}        
\usepackage{graphics} 
\usepackage{epsfig} 
\usepackage{mathptmx} 
\usepackage{times} 
\usepackage{amsmath} 
\usepackage{amssymb}  
\usepackage{algorithm}
\usepackage{algorithmic,subfig}
\usepackage{amssymb,amsmath,psfrag, units, booktabs,algorithm,algorithmic,url}
\usepackage{auto-pst-pdf}
\usepackage{stmaryrd}
\usepackage{caption}
\begin{document}
\begin{frontmatter}
\vspace{-1cm}
\title{On the Practical Design of Tube-enhanced Multi-stage  Nonlinear Model Predictive Control} 

\thanks[footnoteinfo]{\copyright 2022 Sankaranarayanan Subramanian, Yehia Abdelsalam, Sergio Lucia, Sebastian Engell. This work has been accepted to IFAC for publication under a Creative Commons Licence CC-BY-NC-ND.}

\author[First]{Sankaranarayanan Subramanian} 
\author[First]{Yehia Abdelsalam} 
\author[Second]{Sergio Lucia} 
\author[First]{Sebastian Engell}

\address[First]{Process Dynamics and Operations Group, TU Dortmund, Germany (e-mail: \{sankaranarayanan.subramanian, yehia.abdelsalam, sebastian.engell\}@tu-dortmund.de).}
\address[Second]{Process Automation Systems, TU Dortmund, Germany  (e-mail: sergio.lucia@tu-dortmund.de)}

\begin{abstract}                
Tube-enhanced multi-stage nonlinear model predictive control  is a robust control scheme that can handle a wide range of uncertainties with reduced conservatism and manageable computational complexity. In this paper, we elaborate on the {f}{l}exibility of the approach from an application point of view. We discuss the path to making design decisions to implement the novel scheme systematically. We illustrate the critical steps in the design and implementation of the scheme for an industrial example.
\end{abstract}

\begin{keyword}
	Process control, model predictive control, robust control, constrained systems, batch process
\end{keyword}

\end{frontmatter}

\section{Introduction}
Nonlinear Model Predictive Control (NMPC) is an optimization based advanced control strategy that can be used to control nonlinear systems with constraints efficiently \citep{findeisen2002introduction, rawlings2009b}. NMPC uses a nonlinear model to forecast the system's future behavior until a {f}{i}nite horizon and obtains a sequence of control inputs that optimizes a given objective and satisfies the state and input constraints. The control input obtained at the first prediction step is applied to the system, and the optimization problem is solved at the next step in a receding horizon fashion.  The  performance of the NMPC scheme depends critically on the accuracy of the model. The uncertainties present in the model may lead to erroneous forecasts resulting in constraint violation, loss of recursive feasibility, or even closed-loop instability. Hence, to systematically address the presence of uncertainties in the model, robust NMPC schemes are necessary. 

Open-loop min-max model predictive control (MPC) is one of the earliest proposed robust strategies in the context of linear systems \citep{campo1987}. The scheme predicts a sequence of control inputs that minimize the worst-case objective while satisfying the constraints for all realizations of the uncertainty. However, predicting one control input per prediction step may lead to conservative behavior. The closed-loop min-max  MPC \citep{scokaert1998} and the multi-stage MPC formulations \citep{delaPena2005b,lucia2013, lucia2020stability} were proposed to address this conservatism by adapting the control inputs in the predictions for different possible realizations of the uncertainties. This models recourse in the problem formulation and achieves less conservative results. Because of the reduced conservatism, the multi-stage approach has been applied in different case studies  \citep{puschke2018robust,jang2016robust,8374815}.

Unlike the min-max approaches, multi-stage NMPC optimizes the weighted mean of all the scenarios considered in the prediction while accounting for the feedback in the predictions. The drawback of the scheme is that the complexity of the optimization problem grows rapidly when the number of uncertainties and the prediction horizon increase. Tube-based NMPC is another class of robust NMPC approach that provides computationally simpler alternative that generates nominal trajectories optimally and then regulates the processes to converge to or stay close to these (see \citep{ Mayne20111,Rugabotti2011,VILLANUEVA20177175}). In \citep{Mayne20111}, a nominal and an ancillary controller are employed hierarchically to achieve robustness. In the presence of significant uncertainties, tube-based NMPC can be conservative while the complexity remains close to the nominal NMPC controller. 

To improve the trade-off between optimality and complexity compared to the existing robust MPC strategies, Tube-enhanced multi-stage (TEMS) MPC  was proposed for the linear case in \citep{Subramanian2021TEMSlinear} and for the nonlinear case  in \citep{subramanian2018adchem}. The sufficient conditions to achieve closed-loop stability using the TEMS NMPC scheme were derived in \citep{ Subramanian2022}. While it is possible to obtain prior safety and stability guarantees of the closed-loop, it is however very difficult to verify in practice whether the required assumptions are satisfied.

 The TEMS NMPC approach, in its basic form, employs two hierarchical robust multi-stage NMPC controllers. In this contribution, we elaborate on the {f}{l}exibility and the applicability of the TEMS NMPC approach from an application point of view. The following aspects of the design process are explained:
 \begin{enumerate}
\item Construction of the scenario tree for handling different kinds of uncertainties with varying significance.
\item Different formulations of the ancillary controller to achieve robust constraint satisfaction.
\item Constraint tightening of the primary controller.
\item Initialization of the scenario tree for the primary controller.
 \end{enumerate}
 This is followed by the application of the method to an industrially relevant fed-batch polymerization process.

\section{System description and control goals}
The dynamics of the system is described as follows:
\begin{align}\label{eq_true_system}
x^+=f(x,u,d),
\end{align}
where $x\in\mathbb{R}^{n_x}, u\in\mathbb{R}^{n_u},\,d\in \mathbb{D}\subset\mathbb{R}^{n_d}$ denote the states, inputs and model uncertainties (both parametric and additive uncertainties). The nonlinear mapping $f:\mathbb{R}^{n_x }\times \mathbb{R}^{n_u}\times \mathbb{R}^{n_d}\to \mathbb{R}^{n_x}$ denotes the system dynamics.  The uncertainty bound $\mathbb{D}$ is assumed to be compact and can be represented  as $\mathbb{D}=\{d\mid d^L\leq d\leq d^U\}$, where $d^L$ and $d^U$ denote the lower and the upper bounds of the uncertainty. The state constraints are denoted by  $\mathbb{X}$, and the input constraints are denoted by $\mathbb{U}$. The constraint set $\mathbb{X}$ is assumed to be closed and $\mathbb{U}$  is  assumed to be compact. The control goal can be  either to track a setpoint or to economically optimize the performance of the plant while satisfying the state and input constraints for all realizations of the uncertainties at all times. In addition, the controller should not be overly conservative and should be real-time implementable. Here, we assume for simplicity that full-state information is available.
\section{Tube-enhanced multi-stage NMPC}

\tikzstyle{block} = [draw, fill=blue!20, rectangle, 
minimum height=2em, minimum width=4em]
\tikzstyle{sum} = [draw, fill=blue!20, circle, node distance=1cm]
\tikzstyle{input} = [coordinate]
\tikzstyle{output} = [coordinate]
\tikzstyle{pinstyle} = [pin edge={to-,thin,black}]

Tube-enhanced multi-stage NMPC employs two controllers hierarchically to achieve the control goals:  a primary multi-stage NMPC controller and an ancillary multi-stage NMPC controller. The primary controller achieves robustness of the primary system, for which the dynamics are described by
\begin{align}\label{eq_primary}
z^+=f(z,v,\overline{d})
\end{align}
where $z\in \mathbb{R}^{n_x}$ denotes the state, $v\in \mathbb{R}^{n_u}$ denotes the control input, and $\overline{d}\in\overline{\mathbb{D}}\subset \mathbb{R}^{n_{{d}}}$ denotes the uncertainties of the primary system. The nonlinear mapping  $f:\mathbb{R}^{n_x }\times \mathbb{R}^{n_u}\times \mathbb{R}^{n_{d}}\to \mathbb{R}^{n_x}$ is the same as in \eqref{eq_true_system} for all $\overline{d}\in\overline{\mathbb{D}}$. The uncertainty set $\overline{\mathbb{D}}$ is a finite set and is generated by sampling the true uncertainty set $\mathbb{D}$. The predicted trajectories generated by the primary controller are then tracked by the ancillary controller to mitigate the uncertainties that are not considered in the primary system. 

\subsection{Primary controller}
To achieve robustness against the uncertainties that are considered in the primary system described in \eqref{eq_primary},  multi-stage NMPC is employed. The optimization problem formulation for the primary controller is given below:
\begin{subequations} \label{ch4_eq_objective_general}
	\begin{align}\label{ch4_eq:objective}
	\min_{v_k^j \forall (j,k)\in I_{[ 0, N-1] }} \;\;\;\sum_{k=0}^{N-1}\sum_{j=1}^{s^{k}}\omega_k^j \ell(z_k^{j},v_{k}^j) +\sum_{j=1}^{s^N}V_f(z_N^j), 
	\end{align}
	\text{subject to:}
	\begin{align}
	& z_{k+1}^{c(j,r)}=f(z_k^{j},v_k^j,d^{r}),&&\, \forall\, (j,k) \in I_{[ 0, N-1] },\, r\in\{1,\,\dots,\,s\},\label{eq_pc_model}\\
	&\qquad z_k^j\in\mathbb{Z},  v_k^j\in\mathbb{V},&&\, \forall \; (j,k) \in I_{[ 0, N-1] },\label{eq_state_input}\\	
	&\qquad z_N^j\in\mathbb{Z}_f, &&\, \forall \; (j,N) \in I_N,\label{eq_term}\\
	&v_k^j=v_k^l\; \text{if } z_k^{j}=z_k^{l} &&\, \forall \; (j,k), (l,k) \in I_{[ 0, N-1] },\label{eq_on_anti}
	\end{align}
\end{subequations}
where $N$ denotes the prediction horizon, $s$ denotes the number of branches in the scenario tree, $I$ denotes the set of indices $ (j,k)$ in the scenario tree, $I_{[k_1,k_2]}$  denotes the set of indices from the time step $k_1$ until the step $ k_2$, $I_N$ denotes the indices at the terminal nodes, $\ell(z_k^j,v_k^j)$  denotes the stage cost, and $V_f (z_N^j)$ denotes the terminal cost. The weights associated with each node are denoted by $\omega_k^j,\forall(j,k)\in I$.   \eqref{eq_pc_model} denotes the model of the system where the predicted state $z_{k+1}^{c(j,r)}$ is  the child node obtained from its parent state $z_k^j$  for the input $v_k^j$ and the realization of the uncertainty $d_k^{r(j)}\in\overline{\mathbb{D}}$. The state and input constraints are defined in \eqref{eq_state_input}. The terminal constraints are defined in \eqref{eq_term}. The state constraint set is denoted by $\mathbb{Z}$, the input constraint set is denoted by $\mathbb{V}$, and the terminal constraint set is denoted by $\mathbb{Z}_f$. Both $\mathbb{Z}$ and $\mathbb{V}$  are tightened compared to original constraint sets, and $\mathbb{Z}_f\subseteq\mathbb{Z}\subseteq\mathbb{X}$ and $\mathbb{V}\subseteq\mathbb{U}$ hold. The non-anticipativity constraints are defined in \eqref{eq_on_anti}. The realized uncertainty of the primary system $\overline{d}(t)$ for all $t\geq0$ can  be estimated by solving the optimization problem:
\begin{align}\label{eq_min_dist_init}
\overline{d}(t){\in}\arg\min_{\overline{d}\in\mathbb{\overline{D}}} \vert x(t+1)-f(x(t),u(t),\overline{d})\vert,
\end{align}
where $x(t)$ and $x(t+1)$ denote the state measurements at time steps $t$ and $t+1$. The resulting $\overline{d}(t)$ is used to obtain $z(t+1)$ using \eqref{eq_primary}, where $\overline{d}=\overline{d}(t)$.
\subsection{Ancillary controller}
 Robustness against the uncertainties that are not explicitly part of the primary system \eqref{eq_primary} is achieved by the ancillary controller. The ancillary controller tracks the predictions of the primary controller, and the optimization problem formulation is
\begin{subequations} \label{ch4_eq_objective_general_anc}
	\begin{align}\label{eq:objective_anc}
	\min_{u_k^j \forall (j,k)\in I_{[ 0, N-1] }} \;\;\;\sum_{k=0}^{N-1}\sum_{j=1}^{s^{k}}\omega_{a,k}^j \ell_a(x_k^j-{z}_k^{j*},u_k^{j}-{v}_k^{j*}) +\sum_{j=1}^{s^N}V_{fa}(x_N^j), 
	\end{align}
	\text{subject to:}
	\begin{align}
	& x_{k+1}^{c(j,r)}=f(x_k^{j},u_k^j,d^{r}),\,&& \forall (j,k) \in I_{[ 0, N-1] },\, r\in\{1,\dots,s\},\\
	& u_k^j\in\mathbb{U},\, &&\forall  (j,k) \in I_{[ 0, N-1] },\\
	&u_k^j=u_k^l\; \text{if } x_k^{j}=x_k^{l},\, &&\forall  (j,k), (l,k) \in I_{[ 0, N-1] }.\label{ch4_eq_non_anti}
	\end{align}
\end{subequations}
where  the stage cost of the ancillary controller is defined as	$\ell_a(x_k^j-{z}_k^{j*},u_k^{j}-{v}_k^{j*})$ and is tuned to   track the optimal state and input trajectories that were obtained from the primary controller ${z}_k^{j*}$ and ${v}_k^{j*}$  for all $(j,k)\in I_{[ 0, N-1] }$. The terminal cost is denoted as  $V_{fa}(x_N^j)$.  The weights associated with the nodes of the scenario tree of the ancillary controller are denoted by $\omega_{a,k}^j$ for all $(j,k)\in I$. The state constraints are not part of the ancillary controller problem. This enables the recursive feasibility of the ancillary controller by construction. The robust constraint satisfaction is achieved by tuning the stage cost of the ancillary controller. 
At the initial time step $t=0$, the primary controller is initialized with $x(0)$ and $u=v_0^{1*}$ is applied to the system. For all $t\geq 1$, the implementation steps are summarized in Algorithm~\ref{ch4_algorithm}.
\begin{algorithm}
	\caption{ Tube-enhanced multi-stage NMPC algorithm}\label{ch4_algorithm}
	\begin{tabular}[]{p{0.13\columnwidth} p{0.8\columnwidth}}
		\textbf{Require} & All the ingredients of \eqref{ch4_eq_objective_general} and \eqref{ch4_eq_objective_general_anc}.\\
		Step 1&Measure the state $x(t)$. \\
		Step 2&Solve the optimization problem \eqref{eq_min_dist_init} and obtain $\overline{d}(t-1)$ and simulate \eqref{eq_primary} to obtain $z(t)$.\\
		Step 3&Solve \eqref{ch4_eq_objective_general}  using $z_0^1=z(t)$  to get $\{{z}_k^{j*}\}, \,\forall (j,k)\in I_{[ 0, N] }$ and $\{{v}_k^{j*}\},\,\forall (j,k)\in I_{[ 0, N-1] }$.\\
		Step 4&Initialize the ancillary controller with $x(t)$, pass the reference trajectories $\{{z}_k^{j*}\},\,\forall (j,k)\in I_{[ 0, N] }$ and $\{{v}_k^{j*}\},\,\forall (j,k)\in I_{[ 0, N-1] }$ and solve~\eqref{ch4_eq_objective_general_anc}.\\
		Step 5&Apply $u_0^{1*}$ as the input to the plant, set $t=t+1$ and go to step 1 at the next time step.
	\end{tabular}
\end{algorithm}
\\The closed-loop dynamics of the  composite system is given by
\begin{subequations}\label{combined1}
	\begin{align}
	x^+&= f( x,\kappa_{N}(x,z),d),\, d\in\mathbb{D},\label{close_1}\\
	z^+&=f(z,\kappa^p_{N}(z),\overline{d}),\,\overline{d}\in\mathbb{\overline{D}},\label{close_2}
	\end{align}
\end{subequations}
where $\kappa_{N}(x,z)=u_{0}^{1*}$ denotes  the control law of the ancillary controller obtained by solving the optimization problem \eqref{ch4_eq_objective_general_anc}, $\kappa_{N}^p(z,t)=\overline{v}_{0}^{1*}$ denotes the control law for the primary system~\eqref{eq_primary}, {$\overline{d}$ is obtained by solving \eqref{eq_min_dist_init}, and $d\in\mathbb{D}$ denotes the realized uncertainty in the plant. 
	
The differences among the robust NMPC schemes are illustrated in Figure~\ref{fig_comparison}. Multi-stage NMPC \citep{lucia2013} typically requires more branches to handle the same set of uncertainties  compared to  TEMS NMPC and employs a single controller which is initialized with the true state of the system. Tube-based NMPC \citep{Mayne20111} uses two controllers as in the TEMS NMPC scheme but employs only a single scenario. This can lead to stringent constraint tightening in the presence of significant uncertainties and therefore negatively affect the performance. Considering only the significant uncertainties in the TEMS NMPC controllers results in an improved trade-off between optimality and complexity.
\begin{figure}[h]
	\begin{center}
		\includegraphics[width=0.85\columnwidth]{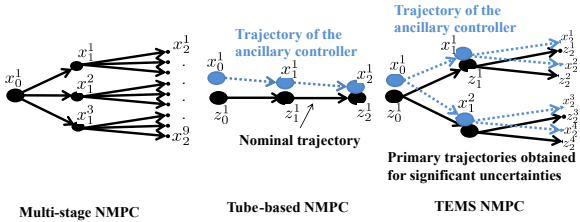}
		\caption{Illustration of multi-stage NMPC, tube-based NMPC, and TEMS NMPC.}  
		\label{fig_comparison}                                 
	\end{center}  
\end{figure}
\section{ Design details}
Before discussing the design steps, we first discuss ways to simplify the controllers.
\subsection{Controller simplifications}\label{sec_simple}
The scenario tree of the primary and ancillary controller problems grows exponentially with respect to the prediction horizon. To avoid the growth in problem complexity, the uncertainty realizations can be assumed to remain constant or a disturbance model can be employed after a certain prediction step called robust horizon $N_R$. Beyond $N_R$, there exists only  one child per parent node. Though there exists no prior guarantees, robust constraint satisfaction and recursive feasibility are often achieved in practice even with $N_R=1$ as demonstrated in \citep{lucia2013,lucia2014e}. 

The second possible simplification is the choice of the ancillary controller. Instead of tracking all the state and input trajectories of the primary controller, only a subset of the trajectories that are critical from control perspective can be tracked. In the simplest possible case, a single trajectory can be tracked. In this case, the ancillary control problem simplifies as follows:
\begin{subequations} \label{ch4_eq_objective_general_anc_sim}
	\begin{align}\label{eq:objective_anc_sim}
	\min_{{u_k} ,\forall k\in \{0,\dots,N-1\}} \;\;\;\sum_{k=0}^{N-1}\ell_a(x_k-{z}_k^{*},u_k-{v}_k^{*}) +V_{fa}(x_N), 
	\end{align}
	\text{subject to:}
	\begin{align}
	& x_{k+1}=f(x_k,u_k,d_{\mathrm{nom}}),&&\quad \forall\, k\in\{0,\dots,N-1\},\\
	& u_k^j\in\mathbb{U},& &\quad \forall\, k\in\{0,\dots,N-1\},
	\end{align}
\end{subequations}
where only the trajectories corresponding to the nominal uncertainty $d_{\mathrm{nom}}$  are tracked. This reduces the complexity of the online implementation of the ancillary controller. Instead of tracking a nominal trajectory, one can even adapt the trajectories to be tracked depending on the realization of the uncertainty \citep{ABDELSALAM202011563}.

\subsection{ Construction of the controllers}\label{sec_const}

The scenario tree is constructed for all the realizations of the uncertainties in the set $\overline{\mathbb {D}}$. The choice of elements of the set $\overline{\mathbb {D}}$ (defined in \eqref{eq_primary}) is the first design decision to make. A heuristic for  scenario tree construction is to consider extreme values of the uncertainties along with the nominal values of the parameter. In \cite{lucia2014e}, it was shown that robust constraint satisfaction was achieved for all $d\in\mathbb{D}$ even for nonlinear systems by considering only the extreme realizations and nominal values of the uncertainties for industrially relevant case studies. However, the problem complexity increases rapidly with the dimension of the uncertainty $n_d$.
 
In  TEMS NMPC, the uncertainties shall be ordered in terms of significance (e.g., based on sensitivity analysis) and  only the most significant uncertainties should be used for the construction of the scenario trees. Let the number of significant uncertainties be denoted as $n_{\overline{d}}$  that is smaller than $n_d$. Let the uncertainty $d$ be defined as $d\triangleq \{d_1,d_2,\dots,d_{n_d}\}$, the elements of which are ordered such that the first $n_{\overline{d}}$ elements are significant. Let $\overline{\mathbb{D}}_c$ be defined as 
 \begin{align}\label{eq_cont_set}
\overline{\mathbb{D}}_c=\bigg\{d\mid& d_i^L\leq d_i\leq d_i^U,\forall i\in\{1,\dots,n_{\overline{d}}\},\nonumber\\&d_i=d_{\mathrm{nom}},\forall i\in\{n_{\overline{d}}+1,\dots,n_d\}\bigg\}.
 \end{align}
 Then $\overline{\mathbb{D}}$ can be constructed by sampling the extreme and nominal values of all $d_i,\, i\in\{1,\dots,n_{\overline{d}}\}$.  Various works in the literature concerning scenario tree generation can also be employed in the context of TEMS NMPC: quadrature-based approximation \citep{leidereiter2014quadrature}, vector quantization approach \citep{goodwin2009vector} or sensitivity-based construction \citep{thombre2021sensitivity}. The scenario tree of the ancillary has the same structure used in the primary controller. However, there exist computationally simpler alternatives as discussed in Section~\ref{sec_simple}.

\subsection{Constraint tightening of the primary controller and tuning the ancillary controller}\label{sec_const_tight}

The constraints of the primary controller should be tightened to provide room for the ancillary controller to satisfy the constraints of the original system. There is an interplay between the stage costs for the ancillary controller and the constraint tightening of the primary controller. Hence during the simulation study, it is recommended to study the closed-loop response as the first step. The stage cost of the ancillary controller can be chosen as a quadratic function as follows:
\begin{align}\label{eq_anc_stg_cost}
\ell_a(x-z,u-v)=(x-z)^TQ(x-z)+(u-v)^TR(u-v),
\end{align}
where $Q$ and $R$ are positive semi-definite matrices. For simpler implementations, $Q$ and $R$ can be chosen as diagonal matrices with non-negative elements. The gains can be chosen to track the critical states and inputs. If the tracking of the ancillary controller is satisfactory, the primary controller constraints can be tightened to account for the magnitude of the constraint violations.

A simple way to tighten the constraints is by simulation studies as proposed in~\cite{Mayne20111}. Let the  state and input constraints  be represented as $g_i(x,u)\leq 0$, $i=\{1,2,\cdots,n_c\}$, where $n_c$ denotes the number of constraints.  The constraints of the primary controller can be represented as $g_i(x,u)\leq -\delta_i$, $i=\{1,2,\dots,n_c\}$. The parameters $\delta_i\geq 0,\,\forall  i=\{1,2,\dots,n_c\}$ associated with  each constraint can be chosen based on the maximum violations observed for the final tuning of the ancillary controller. Once the constraints are tightened,  rigorous constraint satisfaction can be verified by performing extensive  simulation studies for different values of uncertainties and for different initial conditions. 

\subsection {Reinitialization and solution of the controllers}

 Both \eqref{ch4_eq_objective_general} and \eqref{ch4_eq_objective_general_anc}  are  solved recursively. At the initial step $(t=0)$, the primary controller is initialized with the true state (or the initial estimate) of the system. For all $t>0$, the primary controller is always initialized with one of its predicted states. This is because \eqref{eq_min_dist_init} finds the closest uncertainty realization considered in the scenario-tree branches to the measured state. The optimization problem \eqref{ch4_eq_objective_general}  can be solved in previous time steps for all uncertainty realizations considered in the scenario tree. As soon as the measurement is received, the trajectories corresponding to the realized uncertainty can be passed to the ancillary controller.  In some cases, it is possible that the optimization problem \eqref{eq_min_dist_init} results in {f}{l}uctuations of the uncertainty because of various disturbances.  If the disturbance dynamics are available, they can be employed to improve the accuracy of the resulting estimates of $\overline{d}$. An alternative is to include regularization term for all $t\geq 1$ in \eqref{eq_min_dist_init} as follows:
 \begin{align*}
 \overline{d}(t){\in}\arg\min_{\overline{d}\in\mathbb{\overline{D}}} \vert x(t+1)-f(x(t),u(t),\overline{d})\vert+\vert W( \overline{d}(t-1)-\overline{d})\vert,
 \end{align*}
 were $W$ is a weighting matrix. The weighting matrix can be chosen as a diagonal matrix with large values for  time-invariant or slowly time varying parameters and small values for rapidly time-varying parameters. To improve the performance of the closed-loop, the parameters of the primary system can be estimated for all $t>0$ as follows:
  \begin{align}\label{eq_min_dist_init_imp}
 \overline{d}(t){\in}\arg\min_{\overline{d}\in\mathbb{\overline{D}}_c} \vert x(t+1)-f(x(t),u(t),\overline{d})\vert+\vert W( \overline{d}(t-1)-\overline{d})\vert,
 \end{align}
 where $\mathbb{\overline{D}}_c$ contains the closed intervals of the parameters considered in the scenario tree as defined in \eqref{eq_cont_set}. The primary state at the next time step ($z(t+1)$) can be computed from the current values of the state and input of the primary controller and the resulting uncertainty of the problem~\eqref{eq_min_dist_init_imp},  using \eqref{eq_primary}. If \eqref{eq_min_dist_init_imp} is employed, the primary dynamics \eqref{eq_primary} are defined for all $\overline{d}\in\overline{\mathbb{D}}_c$.  When the initialization based on \eqref{eq_min_dist_init_imp} is employed, it is assumed that the primary controller can handle all uncertainties  $\overline{d}\in\mathbb{\overline{D}}_c$.
 The ancillary controller is always initialized with the true state of the system and the solution is computed online. 
\subsection{Tracking vs. economic objectives}

For tracking objectives, a guaranteed stabilizing control law using TEMS NMPC is possible. See \citep{Subramanian2022}  for the necessary  assumptions for prior guarantees.  If the system under investigation is too complex to verify the assumptions, the design steps provided here can help achieve required robust constraint satisfaction and closed-loop stability. Offset-free tracking requires additional steps which are beyond the scope of the paper. The scheme implemented based on the discussed methods  may lead to offsets at steady-state. 

 In the case of economic objectives, the system often operates at the system boundary. Employing multi-stage NMPC, ignoring minor disturbances, may lead to oscillatory input moves (as a response to minor state constraint violations). This oscillation issue is resolved in the TEMS NMPC because the primary controller is affected only by the uncertainties considered in the tree and the ancillary controller does not have state constraints. This avoids aggressive  moves from the ancillary controller if tuned properly. At the same time, implementing tube-based NMPC in the presence of significant uncertainties may lead to conservatism. TEMS NMPC can improve the economic performance because of a less conservative primary controller.

\subsection{Further extensions}

The TEMS NMPC scheme can be easily extended to the output feedback case as well. See \citep{subramanian2018adchem} for details on the same. One can use any efficient estimator to estimate the states of the system and initialize the ancillary controller using the state estimates. The primary controller is implemented the same way as in the full-state feedback case with a more stringent constraint tightening to account also for the estimation errors. 

In the case of time-invariant parameters, suitable parameter estimation algorithms can be used to adapt the scenario tree online. The adaptive TEMS NMPC scheme can yield major performance advantages as shown  in \citep{ABDELSALAM2021}.

\section{Example}
We present the details on the design of TEMS NMPC for the example of a polymerization process that was provided by BASF SE \citep{lucia2014e}. 
There are mass balances of water $m_{\mathrm{W}}$, monomer $m_{\mathrm{A}}$ and polymer $m_{\mathrm{P}}$. There are five energy balances providing the dynamics of temperatures of reactor  ${T}_{\text{R}}$, vessel  ${T}_{\text{S}}$, jacket ${T}_{\text{M}}$, heat-exchanger mixture ${T}_{\text{EK}}$ and coolant at the outlet of the heat exchanger ${T}_{\text{AWT}}$.  The reaction is  exothermic, and the jacket and the heat exchanger are employed to regulate the temperature of the reactor.
The parameter values and the further details of the model can be found in \citep{lucia2014e}. The constraints on the reactor temperature $T_{\text{R}}$ help in achieving the desired quality of the end product, and the adiabatic safety constraint on ${T}_{\text{ad}}$ ensures safe operation of the plant.  The manipulated variables are feed rate of the monomer $\dot{m}_{\text{F}}$, the jacket inlet temperature $T_{\text{M}}^{\text{IN}}$,  and the inlet coolant temperature of the heat exchanger $T_{\text{AWT}}^{\text{IN}}$. The bounds on the state and the control inputs are given in Table 1.

 The reaction rate $k_0$ and the enthalpy of the reaction $\Delta H_{\text{R}}$ are uncertain by $\pm 30\%$ relative to the nominal values. The nominal values of  $\Delta H_{\text{R}}$  and $k_0$ are $-950 $kJ kg$^{-1}$  and $7$, respectively. The bounds are  given by $\Delta H_{\text{R}}\in[-665, -1235]$kJ kg$^{-1}$ and $k_0\in[4.9,9.7]$. In addition, the dynamics of all the states are affected by additive disturbances whose bounds at each discrete time-steps are given by $\pm 0.5$ kg/h for the mass balance of water, $\pm 5$ kg/h for the mass balances of monomer and polymer, and $\pm 0.1^{\circ}$C for the energy balances at discrete time steps. The magnitudes of the uncertainties are different when compared to the simulation study presented in \citep{subramanian2018adchem}.
\begin{table}[htbp]
	\centering
	Table 1. State and input constraints
		\label{table_constraints_states}
		\begin{tabular}{p{0.75cm} p{0.75cm} p{0.75cm} p{2.2cm} p{1.1cm}}
			\hline
			\textbf{State/ input}  & \textbf{Min.} & \textbf{Max.}  & \textbf{Primary controller bounds} &   \textbf{Unit} \\\midrule
			$T_{\text{R}} $  &88.0&92.0   & [88.3, 91.7]&$ ^{\circ} \text{C}$    \\
			$T_{\text{ad}}$ &0 &109 & [1, 108]& $^{\circ} \text{C}$\\
						$\dot{m}_{\text{F}}$                    &0&30000 &[0, 29990]    & $\frac{\text{kg}}{\text{h}}$   \\
			$T_{\text{M}}^{\text{IN}}$ &60&100 &[61, 99]& $^{\circ} \text{C}$  \\
			$T_{\text{AWT}}^{\text{IN}}$    &60&100  &[61, 99]      & $^{\circ} \text{C}$  \\
			\bottomrule
		\end{tabular}
\end{table}
\subsection{Scenario tree construction}
For the construction of the scenario tree, only the parametric uncertainties were considered because they are the most significant ones. Using the heuristics discussed in  Section~\ref{sec_const}, we built the scenario for the extreme and nominal realizations of $\Delta H_{\text{R}}$ and $k_0$, where $\Delta H_{\text{R}}\in \{-950,-655,-1235\}$ and $k_0\in\{7,4.9,9.1\}$. It can be seen that $s=3^2=9$. For the additive uncertainties, the nominal values are considered. The approach was implemented for a robust horizon $N_R=1$. After the first prediction step, all the uncertainties are assumed to be constant. This results in nine scenarios for both the primary controller and the ancillary controller. If we do not classify the uncertainty and employ multi-stage NMPC, it would require $3^{10}=59,049$ scenarios (including the additive uncertainties) for $N_R=1$. The advantage of TEMS in the complexity reduction is obvious in comparison to multi-stage NMPC.
\subsection{Cost function of the primary controller}
The goal is to produce a required amount of polymer as fast as possible. To achieve this goal, the following  stage cost was chosen as in \citep{lucia2014e} for the primary controller. \begin{equation*}\label{2eq_cost_eco}
\begin{aligned}
\ell( z_{k}^j, u_k^j)&=-{m}_{\text{P},k}^{j}+r_{1}(\Delta \dot{m}_{\text{F},k}^j)^2+r_2 (\Delta  T_{\text{M},k}^{\text{IN},j})^2+r_3 (\Delta  T_{\text{AWT},k}^{\text{IN},j})^2,
\end{aligned}
\end{equation*}
where $r_1$, $r_2$ and $r_3$ are tuning parameters that penalize the control moves. The values of the tuning parameters were chosen as $r_1=0.125$, $r_2=4$ and $r_3=0.25$. The terminal cost was chosen as $V_f(z)=0,\,\forall z\in\mathbb{R}^{n_x}$ and the terminal set is chosen as $\mathbb{Z}_f=\mathbb{Z}$. The sampling time was chosen as fifty seconds and the prediction horizon of $N=20$ was chosen. Extensive simulation studies were performed for all uncertainties in $\overline{\mathbb {D}}$ without constraint tightening.
\subsection{Cost function of the ancillary controller}
The stage cost of the ancillary controller is chosen as a quadratic function  defined in \eqref{eq_anc_stg_cost}. The input weight matrix was fixed to identity matrix $R=\mathtt{I}^{3\times 3}$. The tuning of $Q$ was done  considering two aspects: Good tracking of $m_{\text{P}}$ achieves the required goal and  tracking of $T_{\text{R}}$ is required to guarantee robust constraint satisfaction. The constraint $T_{\text{R}}$ remains active throughout the batch and hence, poses a challenge in tuning of $Q$. After several tries, $Q$ was chosen as $\text{diag}(0,0,1,500,0,0,0,0)$. The gain of $500$ was required because of the differences in the absolute values of $m_{\text{P}}$ and $T_{\text{R}}$. If smaller gains were chosen for tracking $T_R$, the constraint violations were relatively large at the end of the batch. The terminal cost of the ancillary controller was chosen as  $V_{fa}(x)=0,\,\forall x\in\mathbb{R}^{n_x}$.
\subsection{Constraint tightening of the primary controller}
After choosing the ancillary stage cost,  simulation studies were performed mainly for the extreme realizations of the uncertainties and the worst-case constraint violations were recorded. The primary controller constraints were tightened based on the procedure given in Section~\ref{sec_const_tight}. The resulting constraint tightening for the primary controller can be found in Table 1.
\subsection{Computation of the solution}
 Orthogonal collocation on finite elements was employed for the discretization of the nonlinear dynamics, and CasADi \citep{andersson2012b} was employed for the automatic generation of first and second-order exact derivatives. The resulting nonlinear programming problem was solved using IPOPT \citep{wachter2006}. For the primary controller, the re-initialization strategy \eqref{eq_min_dist_init_imp} was employed. 
\subsection{Results and comparison}
After all the aspects of the controller had been tuned, a uniform grid of $100$ realizations was chosen in the uncertain parametric space. Each realization was selected for a batch run and $100$ batch runs were simulated. The additive disturbances were varied randomly at every time step during the batch. Selected closed-loop trajectories obtained by applying  TEMS NMPC scheme for the industrial polymerization process are illustrated in Figure \ref{fig_TEMS}. It can be seen that both the quality and safety constraints are satisfied at all times. Also no constraint violations were noted for all $100$ batch runs.

The method is compared with the tube-based NMPC and the multi-stage NMPC approaches. As explained above, considering all the uncertainties in the multi-stage NMPC controller leads to a high computational complexity. Completely ignoring the additive uncertainties resulted in feasibility issues because of violations of the constraint on $T_{\text{R}}$. The additive disturbance on the dynamics of  $T_{\text{R}}$ was therefore considered together with the parametric uncertainties in the scenario tree. For the implementation of multi-stage NMPC, three uncertainties were considered leading to $27$ scenarios are employed. It was very difficult to tune the tube-based NMPC because the nominal trajectory varied considerably compared to the uncertain trajectory. The tube-based NMPC was implemented with very stringent constraint tightening. Even then,  robust constraint satisfaction was not possible for all the batches. 
\begin{figure}[h]
	\begin{center}
		\vspace{-0.6cm}
		\includegraphics[width=0.8\columnwidth]{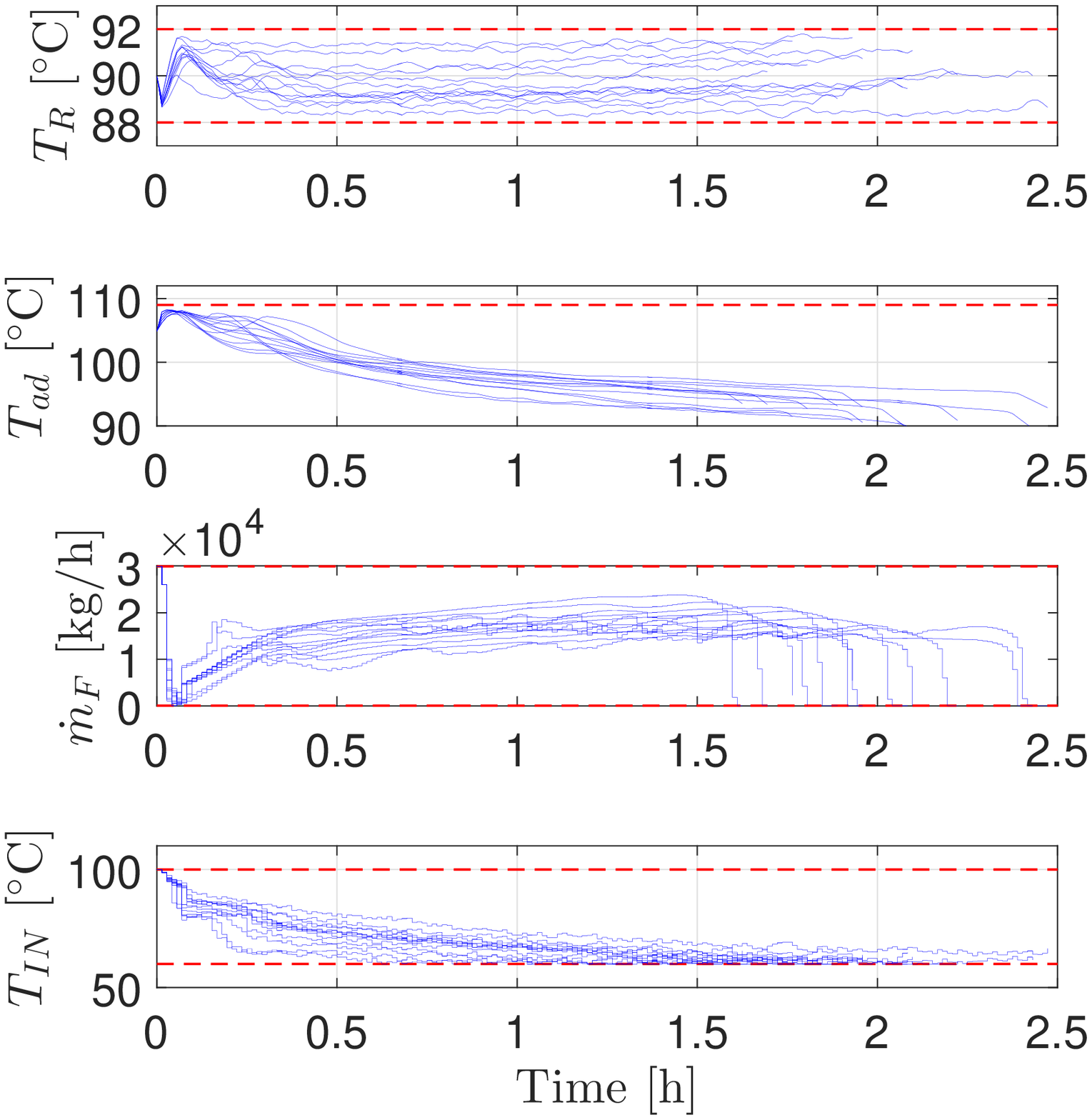} 
		\vspace{-0.5cm} 
		\caption{Trajectories of reactor temperature $T_{\text{R}}$, adiabatic safety temperature $T_{\text{ad}}$, feed rate 	$\dot{m}_{\text{F}}$, jacket inlet temperature $T_{\text{M}}^{\text{IN}}$ for various simulation runs of the TEMS NMPC scheme.  }  
		\label{fig_TEMS}                                 
	\end{center}  
\end{figure}
\begin{table}
	{Table 2. Performance comparison between multi-stage NMPC, tube-based NMPC (Tube)  and TEMS NMPC scheme  concerning batch times, constraint satisfaction and computation times}
	\begin{center}
		\begin{tabular}{p{3.5cm} p{0.9cm} p{0.9cm} p{0.9cm}p{0.9cm}}\label{tab:performance2}
			\textbf{Parameters} & \textbf{MS} &\textbf{Tube}& \textbf{TEMS}\\
			\midrule
			Average batch time [h] 					& 2.1 &3.1& 1.96 \\
			No. of violation of $T_{\text{R}}$ 	& 0&0& 0\\
			No. of. violation of $T_{\text{ad}}$  & 0&14&0\\
			Avg. comp. time per iter. [s] 	& 3.07	&0.07& 0.98 \\
			\bottomrule
		\end{tabular}
	\end{center}
\end{table}

The results are given in Table 2. It can be seen that the TEMS NMPC scheme yields the best average batch time. The average computational complexity was less than one second. It can also be seen that no constraints were violated. Multi-stage NMPC had a higher computational cost. However, it does not result in an improved performance. This is because there were small oscillations in the control moves due to ignored disturbances. Tube-based NMPC on the average leads to 1 hour longer batch times when compared to both multi-stage NMPC and TEMS NMPC.  Tube-based NMPC requires the smallest computation time, but it also resulted in violations of safety constraints in $14$ batches. Overall, it can be concluded that the TEMS NMPC scheme provides a better trade-off between optimality and complexity compared to the other two approaches.
\section{Conclusion}
Tube-enhanced multi-stage NMPC offers  {f}{l}exibility to achieve the desired trade-off between optimality  and complexity. We presented various design details, simplifications, improvements and extensions of the tube-enhanced multi-stage NMPC approach. The scheme was applied to an industrially relevant example and all the design steps were discussed. The resulting scheme showed superior performance when compared to both the multi-stage NMPC and the tube-based NMPC schemes both in terms of performance and computational complexity.
\bibliography{references_All}             
\end{document}